\documentclass[12pt]{elsarticle}
\usepackage{hhline}
\usepackage{longtable}
\usepackage{amsmath}
\usepackage{fancyvrb}
\newfont{\myBbb}{msbm10 scaled 1200}
\newcommand{\mymod}[1]{\ (\bmod\ #1)}

\newcommand{\N}{{\mbox{\myBbb N}}}
\newcommand{\F}{{\mbox{\myBbb F}}}

\newcommand{\be}{\begin{equation}}
\newcommand{\ee}{\end{equation}}
\newcommand{\bea}{\begin{eqnarray}}
\newcommand{\eea}{\end{eqnarray}}

\newcounter{bla}

\journal{Computer Physics Communications}

\begin{document}

\begin{frontmatter}

\title{ PRAND: GPU accelerated parallel random number generation
library: Using most reliable algorithms and applying parallelism
of modern GPUs and CPUs.}

\author[ITP,MIPT]{L.Yu. Barash}
\ead{barash@itp.ac.ru}
\author[ITP,HSE]{L.N. Shchur}
\address[ITP]{Landau Institute for Theoretical Physics, 142432 Chernogolovka, Russia}
\address[MIPT]{Moscow Institute of Physics and Technology, 141700 Moscow, Russia}
\address[HSE]{National Research University Higher School of Economics, 
101000 Moscow, Russia}

\begin{abstract}
The library PRAND for pseudorandom number generation for modern CPUs and GPUs is presented.
It contains both single-threaded and multi-threaded realizations of a number of modern
and most reliable generators recently proposed and studied in~\cite{EPL2011,MT19937,MRG32K3A,LFSR113,CatMaps2006}
and the efficient SIMD realizations proposed in~\cite{RNGSSELIB1}. One of the useful features for using PRAND 
in parallel simulations is the ability to initialize up to $10^{19}$ independent streams.  
Using massive parallelism of modern GPUs and SIMD parallelism of modern CPUs 
substantially improves performance of the generators.
\end{abstract}

\end{frontmatter}

{\bf PROGRAM SUMMARY}

\begin{small}
\noindent
{\em Journal Reference: } Comput. Phys. Commun. 185 (2014) 1343-1353 \\
{\em Program Title:} PRAND \\
{\em Catalogue identifier: } AESB\_v1\_0 \\
{\em Program summary URL: } http://cpc.cs.qub.ac.uk/summaries/AESB\_v1\_0.html \\
{\em Program obtainable from: } CPC Program Library, Queen's University, Belfast, N.~Ireland \\
{\em Licensing provisions:} Standard CPC licence, http://cpc.cs.qub.ac.uk/licence/licence.html  \\
{\em No. of lines in distributed program, including test data, etc.:} 45979 \\
{\em No. of bytes in distributed program, including test data, etc.:} 23953564 \\
{\em Distribution format: } tar.gz \\
{\em Programming language:} Cuda C, Fortran                    \\
{\em Computer:} PC, workstation, laptop, or server with NVIDIA GPU (tested on Tesla X2070, Fermi C2050, GeForce GT540M) and with Intel or AMD processor.\\
{\em Operating system:} Linux with CUDA version 5.0 or later. Should also run on MacOs, Windows, or UNIX  \\
{\em RAM:} 4 Mbytes   \\
{\em Supplementary material:}                                 
{\em Classification:} 4.13 Statistical Methods   \\
{\em External routines/libraries:}                            \\
{\em Subprograms used:}                                       \\
{\em Nature of problem:}  Any calculation requiring uniform pseudorandom
number generator, in particular, Monte Carlo calculations. Any calculation or simulation requiring 
uncorrelated parallel streams of uniform pseudorandom numbers. \\
   \\
{\em Solution method:}
The library contains realization of a number of modern and reliable generators:
\verb#MT19937#, \verb#MRG32K3A# and \verb#LFSR113#.
Also new realizations of the method based on parallel evolution of an ensemble
of transformations of two-dimensional torus are included in the library:
\verb#GM19#, \verb#GM29#, \verb#GM31#, \verb#GM61#,
\verb#GM55#, \verb#GQ58.1#, \verb#GQ58.3# and \verb#GQ58.4#.
The library contains: single-threaded and multi-threaded realizations for GPU,
single-threaded realizations for CPU, realizations for CPU based on SSE command set.
Also, the library contains the abilities to jump ahead
inside RNG sequence and to initialize independent random number streams
with block splitting method for each of the RNGs.
\\
   \\
{\em Restrictions:}
Nvidia Cuda Toolkit version 5.0 or later should
be installed. To use GPU realizations, Nvidia GPU supporting CUDA
and the corresponding Nvidia driver should be installed.
For SSE realizations of the generators,
Intel or AMD CPU supporting SSE2 command set is required.
In order to use the SSE realization of \verb#LFSR113#, CPU must
support SSE4 command set. \\
   \\
{\em Unusual features:}\\
   \\
{\em Additional comments:}
A version of this program, which only contains the realizations for CPUs, is held in the Library 
as Catalog Id., AEIT\_v2\_0 (RNGSSELIB). It does not require a GPU device or CUDA compiler.\\
   \\
{\em Running time:} The tests and the examples included in the package
take less or about one minute to run.
Running time is analyzed in detail in Sec.~\ref{PerformanceSec} of the paper.\\
   \\

\end{small}

\section{Introduction}

Pseudorandom number generation with good statistical
properties is an important component for Monte Carlo
simulations widely used in physics and material
science~\cite{BinderHeerman}.
Pseudorandom number generators (RNGs) used for this purpose
are tested with many hundreds of statistical tests, and some of the RNGs
demonstrate very good results~\cite{RNGSSELIB1,Lecuyer2007}.
However, employing modern supercomputers for
efficient Monte Carlo calculations is limited due to lack of
software which allows to generate large
number of parallel uncorrelated streams of high quality pseudorandom
numbers. Libraries for Monte Carlo simulations should
support modern graphical processing units (GPUs),
which represent generally applicable computational
engines allowing massively parallel computations.
Such software also should support Streaming SIMD Extensions (SSE) technology
allowing to accelerate computations, which is introduced in
Intel Pentium 4 processors in 2001, and is supported by all
modern Intel and AMD processors.

In this paper we present a library of modern and most
reliable RNGs known today for modern CPUs and GPUs.
Namely, the library contains realization of a number of modern and reliable generators:
\verb#MT19937#~\cite{MT19937}, \verb#MRG32K3A#~\cite{MRG32K3A}, \verb#LFSR113#~\cite{LFSR113},
\verb#GM19#, \verb#GM31#, \verb#GM61#~\cite{CatMaps2006,RNGSSELIB1}, and
\verb#GM29#, \verb#GM55#, \verb#GQ58.1#, \verb#GQ58.3#, \verb#GQ58.4#~\cite{EPL2011,Springer2012}.
The library contains: single-threaded and multi-threaded realizations for GPU,
single-threaded realizations for CPU, realizations for CPU based on SSE command set.
Using massive parallelism of modern GPUs and SIMD parallelism of modern CPUs
allows to substantially improve performance of the generators.

In order to carry out a Monte Carlo calculation on a parallel system,
it is necessary to generate large number of sequences
of pseudorandom numbers in parallel.
Block splitting is one of the techniques
that allow to initialize large number of high quality
independent random number streams from a single RNG,
provided that there is the ability to jump ahead
inside RNG sequence (see Sec.~\ref{StreamsSec}).
PRAND library contains the abilities to jump ahead
inside RNG sequence and to initialize large number
of independent random number streams
with block splitting method for each of the RNGs.

\section{Software for random number generation}
\label{softwareSec}

In this section we analyze and compare performance capabilities of
existing software for random number generation.

\subsection{GNU Scientific Library}

The GNU Scientific Library~\cite{GSL} contains realizations for the following random
number generators: borosh13, coveyou, cmrg, fishman18,
fishman20, fishman2x, gfsr4, knuthran, knuthran2,
lecuyer21, minstd, mrg, mt19937, mt19937\_1999,
mt19937\_1998, r250, ran0, ran1, ran2, ran3,
rand, rand48, random128\_bsd, random128\_glibc2, random128\_libc5,
random256\_bsd, random256\_glibc2, random256\_libc5, random32\_bsd,
random32\_glibc2, random32\_libc5, random64\_bsd, random64\_glibc2,
random64\_libc5, random8\_bsd, random8\_glibc2, random8\_libc5,
random\_bsd, random\_glibc2, random\_libc5, random\_libc5,
randu, ranf, ranlux, ranlux389, ranlxd1, ranlxd2,
ranlxs0, ranlxs1, ranlxs2, ranmar, slatec, taus,
taus2, transputer, tt800, uni, uni32, vax,
waterman14, zuf.

Except for several versions of Mersenne Twister,
the generators listed above are, actually,
very old classical RNGs (linear congruential, GFSR, Lagged
Fibonacci, etc.). As follows from the analysis in Section~\ref{PropGen},
such generators should not be used for Monte Carlo
simulations on modern supercomputers.
The library supports only standard realizations in C language for CPU.
Both parallel streams of
pseudorandom numbers and speeding up with SSE or GPU, are not supported.

\subsection{Intel Math Kernel Library}

Vector Statistical Library is a part of Intel MKL Library~\cite{MKL} and contains
realizations for the following RNGs:
MCG31m1 (linear congruential generator), R250 (shift register sequence~\cite{Kirkpartrick}),
MRG32K3A, MCG59 (linear congruential), WH (combination of four linear congruential),
MT19937, MT2203 (generator of the same type as Mersenne Twister~\cite{MT19937,MT2203}),
SFMT19937 (generator of the same type as Mersenne Twister, speeded up with SSE).

Therefore, the library contains six different RNGs and their variations.
These are CPU realizations, speeding up with SSE is supported.
In order to generate parallel streams of pseudorandom numbers,
the library contains 273 versions of WH and 6024 versions of MT2203.
In particular, the authors of MT2203 conjecture that
sequences based on linear recurrences are statistically independent
if the characteristic polynomials are relatively prime to each other,
without any theoretical support. Thus, we do not recommend Intel Math Kernel Library
for generating parallel streams of random numbers:
first of all, it supports initialization of only relatively small number
of parallel sequences,  secondly, in this case parallel streams of pseudorandom
numbers are generated with parametrization method, which does not have
sufficient theoretical support (see Section~\ref{StreamsSec}) and correlations between streams 
could not be excluded.

\subsection{RNGSSELIB}

Our earlier package RNGSSELIB~\cite{RNGSSELIB1} contains modern and
reliable generators: MT19937, MRG32K3A, LFSR113, GM19, GM31, GM61.
The library supports speedup with SSE. The performance of new realizations
exceeds those of Intel Math Kernel Library. The comparison of
performance and detailed statistical testing are presented in~\cite{RNGSSELIB1}.
In the new version of RNGSSELIB~\cite{RNGSSELIB2} we have added new reliable generators GM29, GM55, GQ58.1,
GQ58.3, GQ58.4. The updated library is compatible with Fortran and
includes examples of using  RNGSSELIB in Fortran.
Also the updated library provides abilities to jump ahead inside RNG sequence and to initialize up to $10^{19}$
independent random number streams with block splitting method for each of the RNGs.
The algorithms for jumping ahead and generating independent streams used
in our updated version of RNGSSELIB are essentially the same as those
elaborated in the present paper for PRAND library.
They are described in detail in Sec.~\ref{StreamsSec}.

\subsection{The Scalable Parallel Random Number Generators Library (SPRNG)}

The SPRNG library~\cite{SPRNG}  includes CPU realizations of the following RNGs:
LCG48 (linear congruential), LFG (Lagged Fibonacci Generator),
LCG64 (linear congruential), CMRG (combined MRG-generator),
MLFG (multiplicative Lagged Fibonacci Generator),
PMLCG (linear congruential).

In a recent paper~\cite{GASPRNG} GASPRNG package with
GPU realizations is presented: the package contains
realizations for the same RNGs as SPRNG library.
GASPRNG is a GPU accelerated implementation of the SPRNG
library, so it produce exactly the same sequences
as SPRNG. Most of the RNGs from SPRNG are old classical
random number generators. Their properties are presented
in Table~\ref{rngprop} in Section~\ref{PropGen} and
can be compared there with those of other generators.
SPRNG and GASPRNG packages
support generation of parallel streams of pseudorandom numbers
with parametrization method, which does not have sufficient
theoretical support (see Section~\ref{StreamsSec}).

\subsection{Tina's Random number generator library (TRNG)}

The TRNG library~\cite{TRNG}  contains CPU realizations of the following
RNGs: lcg64, lcg64\_shift, mrg\_, mrg\_s, yarn\_, yarn\_s,
lagfib\_xor, lagfib\_plus, mt19937, mt19937\_64.
In other words, TRNG includes classical linear congruential
generators, MRG generators, Lagged Fibonacci generators,
and Mersenne Twister.
Only six generators from TRNG (lcg64, lcg64\_shift, mrg\_,
mrg\_s, yarn\_, yarn\_s) support GPU and generation of
parallel pseudorandom streams.

\subsection{NAG Numerical Routines for GPUs}

NAG Numerical routines for GPUs~\cite{NAG} are a set of routines which
appeared at the end of 2011. They include the ability to generate
parallel streams for MRG32K3A and MT19937. Source codes of the routines
are unavailable. 

According to~\cite{GPUGemsArticle}, the coefficients
for MT19937 are calculated in NAG Numerical routines for GPUs
during generation of parallel streams.
PRAND realization is different and coefficients are calculated
in advance (see Sec.~\ref{MTJumpAhead}).
Performance tests for NAG routines are presented in
Section~\ref{PerformanceSec}.

\subsection{Nvidia cuRand library}

Nvidia cuRand is a part of Nvidia CUDA Toolkit~\cite{cuRand}, and supports
the generators MTGP Mersenne Twister~\cite{MTGP} and MRG32K3A since
version 4.1 which appeared in February 2012.
cuRand also supports the XORWOW generator.
CuRand supports generation of parallel streams of pseudorandom numbers
with parametrization method for MTGP Mersenne Twister,
which does not have sufficient theoretical support.
In particular, MTGP Mersenne Twister use
sequences based on linear recurrences based on
distinct irreducible characteristic polynomials
in order to create independent streams of pseudorandom numbers.
The property that the generators have distinct
irreducible characteristic polynomials, generally,
does not automatically result in the absence
of correlations in a merged sequence.
The source codes for cuRand library are unavailable.
Performance tests for cuRand library are presented in
Section~\ref{PerformanceSec}.

\subsection{Other recent developments}

The single-stream realization for ATI Graphics Processing Units
of several RNGs, including MT19937, can be found in~\cite{Demchik}.

XORSHIFT generators, which were initially suggested by Marsaglia~\cite{Marsaglia},
are fast and are able to have very large period lengths. They were
later studied in detail in~\cite{Panneton} using the equidistribution 
criterion and found to be not reliable in their original version with
three xorshift operations. 
Recently a new version of the XORSHIFT generator was suggested 
by Manssen et.al. in~\cite{Manssen}, which has a dramatically increased word size. 
The particular parameters of the new RNG
were found using the method and software of Brent described in~\cite{BrentXOR}.
Thus, the XORSHIFT generator suggested in~\cite{Manssen} is definitely
much better than the original Marsaglia versions, while its equidistribution
properties still need to be further analyzed.

Other examples of recently proposed generators, which are not yet included 
in any software package, are so-called counter-based generators.
They originate from cryptographic techniques which have been significantly 
simplified (reducing their cryptographic strength) in order 
to achieve high performance~\cite{Manssen,Salmon}.
Namely, keyed bijections from symmetric-key cryptosystems are used,
which are highly nonlinear transformations.
While the method allows the generation of a large number of parallel streams
using the parametrization method, generation of parallel streams
with this method needs to be further analyzed.
Some other examples of recently proposed methods are reviewed 
in~\cite{Manssen,Howes,Nandapalen}.

\section{ Results of statistical testing and other properties of RNGs.}
\label{PropGen}

Table~\ref{rngprop} shows the properties of the RNGs which are supported by the software
discussed in Section~\ref{softwareSec}.
The properties include the speed of generation, logarithm of period of an RNG,
dimension of approximate equidistribution (see also Sec.~\ref{PropPRAND}),
and the results of statistical testing.
They were tested with a computer based on Intel Xeon 5160, 3 Ghz, 4Mb Cache.
For statistical testing we applied the SmallCrush,
Crush and BigCrush batteries of tests taken from the
TestU01 package~\cite{TestU01}.
The Smallcrush test battery contains 15 tests which
altogether use about $2.3\cdot 10^8$ pseudorandom numbers, 
the Crush test battery contains 144 tests which use about $3.4\cdot 10^{10}$
pseudorandom numbers, and the Bigcrush test battery
contains 160 tests which use about $3.6\cdot 10^{11}$ pseudorandom numbers.
For each battery of tests, Table~\ref{rngprop} displays
the number of statistical tests with p-values outside the
interval $[10^{-10},1-10^{-10}]$. In other words, the number of failed
tests for a given battery is displayed.
If the number of statistical tests with p-values outside the
interval $[10^{-3},1-10^{-3}]$ differs from the number of statistical tests with p-values outside the
interval $[10^{-10},1-10^{-10}]$, then it is displayed inside the parenthesis.
Statistical testing with TestU01 was previously applied to a number of generators
in~\cite{Lecuyer2007}.

\setlength{\LTcapwidth}{\textwidth}
\begin{longtable}{|l|l|c|c|c|c|c|c|}
\caption[Properties of the RNGs from different existing libraries]{Properties
of the RNGs from different existing libraries}
\label{rngprop}\\
\hline
\parbox{.10\textwidth}{Library} &
\parbox{.20\textwidth}{Generator}&
\parbox{.10\textwidth}{Speed of generation, million/sec} &
\parbox{.10\textwidth}{$\log_{10}(T)$, where $T$ is period length} &
\parbox{.10\textwidth}{Dimen\-sion of approx. equidistribution } &
\parbox{.10\textwidth}{Small\-crush (15 tests)} &
\parbox{.10\textwidth}{Crush (144 tests)} &
\parbox{.10\textwidth}{Bigcrush (160 tests)} \\
\hline
{GSL} & b{orosh13} & 2{43.9} & 9{.6} & $-$  & 1{2} & 1{12(115)} & 1{06(109)}\\
\cline{2-8}
 & c{oveyou} & 2{12.7} & 9 & $-$  & 1{2} & 9{0(97)} & 9{7(106)}\\
\cline{2-8}
 & c{mrg} & 2{8.2} & 5{5.7} & $-$  & 0 & 0{(1)} & 0\\
\cline{2-8}
 & f{ishman18} & 6{.8} & 9{.3} & $-$  & 2 & 3{8(41)} & 5{7(65)}\\
\cline{2-8}
 & f{ishman20} & 8{2.0} & 9{.3} & $-$  & 3 & 3{6(41)} & 6{0(66)}\\
\cline{2-8}
 & f{ishman2x} & 7{2.5} & 1{6.9} & $-$  & 0 & 1{(2)} & 1{(4)}\\
\cline{2-8}
 & g{fsr4} & 1{45.7} & 2{916.7} & $-$  & 0 & 2{(3)} & 2\\
\cline{2-8}
 & k{nuthran} & 1{17.6} & 3{8.8} & $-$  & 2 & 9 & 1{7(19)}\\
\cline{2-8}
 & k{nuthran2} & 3{.57} & 1{8.7} & $-$  & 0 & 1{(2)} & 1{(3)}\\
\cline{2-8}
 & l{ecuyer21} & 8{1.7} & 9{.3} & $-$  & 3 & 4{1(47)} & 5{4(62)}\\
\cline{2-8}
 & m{instd} & 8{4.3} & 9{.3} & $-$  & 3 & 4{2(51)} & 5{7(67)}\\
\cline{2-8}
 & m{rg} & 5{3.5} & 4{6.7} & $-$  & 0 & 0 & 0\\
\cline{2-8}
 & m{t19937} & 1{13.4} & 6{001.6} & 6{23} & 0 & 2 & 2\\
\cline{2-8}
 & m{t19937\_1999} & 1{13.4} & 6{001.6} & 6{23} & 0 & 2 & 2\\
\cline{2-8}
 & m{t19937\_1998} & 1{13.4} & 6{001.6} & 6{23} & 0 & 2 & 2{(3)}\\
\cline{2-8}
 & r{250} & 1{70.4} & 7{5.3} & $-$  & 1 & 8{(9)} & 1{5(19)}\\
\cline{2-8}
 & r{an0} & 8{4.2} & 9{.3} & $-$  & 3 & 4{3(51)} & 5{7(64)}\\
\cline{2-8}
 & r{an1} & 9{2.2} & 9{.3} & $-$  & 5 & 8{6(88)} & 5{7(60)}\\
\cline{2-8}
 & r{an2} & 6{2.6} & 1{8.4} & $-$  & 5 & 8{5(87)} & 5{5(56)}\\
\cline{2-8}
 & r{an3} & 8{5.3} & 2{5.6} & $-$  & 0{(1)} & 1{2} & 1{5(20)}\\
\cline{2-8}
 & r{and} & 2{27.8} & 9{.3} & $-$  & 1{0(12)} & 1{07(110)} & 1{00(104)}\\
\cline{2-8}
 & r{and48} & 1{4.9} & 1{4.4} & $-$  & 4{(5)} & 2{1(23)} & 4{7(50)}\\
\cline{2-8}
 & r{andom32} & 1{73.2} & 1{1.1} & $-$  & 8 & 9{8(103)} & 1{12(116)}\\
\cline{2-8}
 & r{andom64} & 1{75.3} & 1{3.5} & $-$  & 4 & 6{3(68)} & 6{8(71)}\\
\cline{2-8}
 & r{andom128} & 1{75.9} & 1{8.4} & $-$  & 1{(3)} & 1{2(14)} & 2{0(25)}\\
\cline{2-8}
 & r{andom256} & 1{47.0} & 2{8.0} & $-$  & 1{(2)} & 8 & 1{1(12)}\\
\cline{2-8}
 & r{andu} & 2{42.1} & 8{.7} & $-$  & 1{4(15)} & 1{25(132)} & 1{30(132)}\\
\cline{2-8}
 & r{anf} & 1{4.5} & 1{3.8} & $-$  & 5 & 2{5(28)} & 5{4(60)}\\
\cline{2-8}
 & r{anlux} & 1{4.3} & 1{70.7} & $-$  & 1{2} & 9{1(92)} & 5{9(62)}\\
\cline{2-8}
 & r{anlux389} & 8{.5} & 1{70.7} & $-$  & 1{2} & 9{1(92)} & 60(63) \\
\cline{2-8}
 & r{anlxs0} & 3{0.1} & 170.7  & $-$  & 1{2} & 9{1} & 5{9(61)}\\
\cline{2-8}
 & r{anlxs1} & 2{0.1} & 170.7  & $-$  & 1{2} & 9{0(92)} & 5{8(61)}\\
\cline{2-8}
 & r{anlxs2} & 1{1.8} & 170.7  & $-$  & 1{2} & 9{1(92)} & 5{8(62)}\\
\cline{2-8}
 & r{anlxd1} & 1{2.0} & 170.7  & $-$  & 0 & 0 & 0\\
\cline{2-8}
 & r{anlxd2} & 6{.5} & 170.7  & $-$  & 0 & 0 & 0 \\
\cline{2-8}
 & r{anmar} & 8{8.2} & 1{70.7} & $-$  & 1{2} & 9{1(92)} & 5{8(62)}\\
\cline{2-8}
 & s{latec} & 9{5.4} & 6{.6} & $-$  & 1{4} & 1{31} & 1{44(146)}\\
\cline{2-8}
 & t{aus} & 1{27.1} & 26.5  & $-$  & 0 & 6 & 6\\
\cline{2-8}
 & t{aus2} & 1{26.9} & 26.5  & $-$  & 0 & 6 & 6\\
\cline{2-8}
 & t{ransputer} & 2{43.9} & 9{.6} & $-$  & 1{0(14)} & 1{11(114)} & 1{07(112)}\\
\cline{2-8}
 & t{t800} & 1{36.2} & 2{40.8} & $-$  & 0 & 4{(7)} & 6\\
\cline{2-8}
 & u{ni32} & 8{5.1} & 6{.6} & $-$  & 2{(3)} & 2{4(25)} & 3{0(36)}\\
\cline{2-8}
 & v{ax} & 2{26.5} & 9{.6} & $-$  & 1{3(14)} & 1{07(110)} & 9{7(100)}\\
\cline{2-8}
 & w{aterman14} & 2{42.4} & 9{.6} & $-$  & 1{2} & 1{11(117)} & 1{11(117)}\\
\cline{2-8}
 & z{uf} & 7{4.2} & 1{92.1} & $-$  & 1{2} & 9{2(93)} & 6{0(63)}\\
\hline
Intel& m{cg31m1} & 5{3.4} & 9{.3} & $-$  & 2 & 4{1(42)} & 56(64) \\
\cline{2-8}
MKL& r{250} & 5{4.3} & 7{5.3} & $-$  & 1 & 8 & 1{4}\\
\cline{2-8}
 & m{rg32k3a} & 2{8.2} & 5{7.5} & 4{5} & 0 & 0 & 0\\
\cline{2-8}
 & m{cg59} & 3{7.4} & 1{7.2} & $-$  & 1 & 1{0} & 1{7}\\
\cline{2-8}
 & w{h} & 7{.9} & 1{2.9} & $-$  & 1 & 1{2} & 2{2}\\
\cline{2-8}
 & m{t19937} & 2{6.3} & 6{001.6} & 6{23} & 0 & 2 & 2\\
\cline{2-8}
 & m{t2203} & 2{5.6} & 6{63.2} & 6{8} & 0 & 2 & 4 \\
\hline
RNG- & m{t19937} & 2{16.0} & 6{001.6} & 6{23} & 0 & 2 & 2\\
\cline{2-8}
SSE- & m{rg32k3a} & 1{51.7} & 5{7.5} & 4{5} & 0 & 0 & 0\\
\cline{2-8}
LIB & l{fsr113} & 1{50.4} & 3{4} & 3{0} & 0 & 6 & 6\\
\cline{2-8}
 & g{m19} & 3{1.6} & 1{1.4} & 1{29} & 0 & 0 & 0\\
\cline{2-8}
 & g{m31} & 2{2.1} & 1{8.7} & 2{10} & 0 & 0 & 0\\
\cline{2-8}
 & g{m61} & 5{.5} & 3{6.7} & 4{15} & 0 & 0 & 0\\
\cline{2-8}
 & g{m29} & 4{8.3} & 1{7.4} & 2{00} & 0 & 0 & 0\\
\cline{2-8}
 & g{m55} & 3{9.3} & 3{0.7} & 3{50} & 0 & 0 & 0\\
\cline{2-8}
 & g{q58.1} & 1{1.4} & 1{7.4} & 2{00} & 0 & 0 & 0\\
\cline{2-8}
 & g{q58.3} & 2{8.3} & 1{7.4} & 2{00} & 0 & 0 & 0\\
\cline{2-8}
 & g{q58.4} & 4{2.0} & 1{7.4} & 2{00} & 0 & 0 & 0\\
\hline
S{PRNG} & l{cg48} & 1{63.1} & 1{4.4} & $-$  & 4{(5)} & 2{0(25)} & 4{9}\\
\cline{2-8}
 & l{fg} & 9{9.7} & 2{4.4} & $-$  & 0 & 0{(1)} & 0\\
\cline{2-8}
 & l{cg64} & 1{19.3} & 1{9.3} & $-$  & 0 & 5 & 7{(10)}\\
\cline{2-8}
 & c{mrg} & 8{8.1} & 6{5.9} & $-$  & 0 & 0 & 0{(2)}\\
\cline{2-8}
 & m{lfg} & 8{2.6} & 3{94.3} & $-$  & 0 & 0 & 0\\
\cline{2-8}
 & p{mlcg} & 5{9.2} & 1{8.4} & $-$  & 0 & 3{(5)} & 3{(5)}\\
\hline
T{RNG} & l{cg64} & 1{1.6} & 1{9.3} & $-$  & 0 & 5{(6)} & 7{(9)}\\
\cline{2-8}
 & l{cg64\_shift} & 1{0.5} & 1{9.3} & $-$  & 0 & 0 & 0\\
\cline{2-8}
 & m{rg5} & 9{.1} & 4{6.7} & $-$  & 0 & 0{(2)} & 0\\
\cline{2-8}
 & m{rg5s} & 7{.8} & 4{6.7} & $-$  & 0 & 0 & 0\\
\cline{2-8}
 & y{arn} & 7{.2} & 4{6.7} & $-$  & 0 & 0 & 0{(2)}\\
\cline{2-8}
 & y{arn\_s} & 6{.5} & 4{6.7} & $-$  & 0 & 0{(1)} & 0{(1)}\\
\cline{2-8}
 & l{agfib\_xor} & 1{1.3} & 2{962.7} & $-$  & 0 & 2 & 2\\
\cline{2-8}
 & l{agfib\_plus} & 1{0.6} & 2{962.7} & $-$  & 0 & 0 & 0{(1)}\\
\cline{2-8}
 & m{t19937} & 9{.4} & 6{001.6} & 6{23} & 0 & 2{(3)} & 2\\
\cline{2-8}
 & m{t19937\_64} & 1{0.1} & 6{001.6} & 3{12} & 0 & 2 & 2\\
\hline
 & xorwow & 347.1 & 38.5 & $-$ & 1 & 9(11) & 13(15) \\
\hline
\end{longtable}

\section{Properties of RNGs in PRAND library}
\label{PropPRAND}

The PRAND library contains realization of a number of modern and reliable generators:
MT19937~\cite{MT19937,RNGSSELIB1}, MRG32K3a~\cite{MRG32K3A,RNGSSELIB1} and LFSR113~\cite{LFSR113,RNGSSELIB1}.
Also the PRAND library contains realizations of the method based on parallel evolution of an ensemble
of transformations of two-dimensional torus:
GM19, GM31 and GM61 are constructed and analyzed in detail in~\cite{CatMaps2006,RNGSSELIB1};
GM29, GM55, GQ58.1, GQ58.3 and GQ58.4 are constructed and analyzed in detail in~\cite{EPL2011,Springer2012}.

Table~\ref{ParametersTable} shows the parameters and properties of the generators under discussion.
In particular, the last column shows the dimension of equidistribution
for the generators. High-dimensional uniformity and the corresponding
equidistribution property is one of the most important properties
characterizing the quality of pseudorandom sequences of 
numbers~\cite{Equidistr1,Equidistr2,Equidistr3,Equidistr4,PreviousToLFSR113}.
In order for the exact $n$-dimensional equidistribution up to $v$-bit accuracy
to be satisfied for a generator,
a natural probability measure should be introduced for its output sequences, and
it should be proved for this measure that all output $v$-bit sequences of length $n$ are equiprobable
(see, e.g.,~\cite{EPL2011,Springer2012} for details).
If the exact equidistribution does not hold, one may consider
the set $A_n=\{v^n p(0),v^n p(1),\dots,v^n p(v^n-1)\}$, where $p(i)$ is the probability
of a particular $v$-bit sequence of length $n$ (there are exactly $v^n$ such sequences).
If the variance $\sigma(A_n)$ is much smaller than the mean value $\langle A_n\rangle =1$,
we will say that approximate $n$-dimensional equidistribution is satisfied
for a generator.
MT19937 satisfies exact $623$-dimensional equidistribution~\cite{MT19937}.
MRG32K3A satisfies approximate $45$-dimensional equidistribution~\cite{MRG32K3A}.
LFSR113 satisfies exact $n$-dimensional equidistribution, where $n\approx 30$
~\cite{PreviousToLFSR113,LFSR113}.
The generators GQ58.1, GQ58.3, GQ58.3 satisfy exact $29$-dimensional equidistribution,
GM55 satisfies exact $4$-dimensional equidistribution; also the generators
of this type satisfy approximate $n$-dimensional equidistribution, $n\approx 6.8\log_2 p$,
where $g=p\cdot 2^t$ and $p$ is a prime integer~\cite{EPL2011,Springer2012}.

\begin{table}[p]
\caption{Parameters of the generators}
\label{ParametersTable}
\begin{tabular}{|l|c|c|c|c|c|c|}
\hline
Generator & $k$ & $q$ & $g$ & $v$ & Period & Dimension of approx.\\
                              &&&&& length & equidistribution \\
\hline
GM19        & $15$ & $28$ & $2^{19}-1$ & $1$ & $2.7\cdot 10^{11}$  & $129$\\
GM29        & $4  $ & $2  $ & $2^{29}-3        $ & $1$ & $ 2.8\cdot 10^{17}   $ & $200$ \\
GM31        & $11$ & $14$ & $2^{31}-1$ & $1$ & $4.6\cdot 10^{18}$  & $210$ \\
GM55        & \hspace{-1mm}$256$\hspace{-1mm} & \hspace{-1mm}$176$\hspace{-1mm} & $16(2^{51}-129)  $ & $4$ & $\geq 5.1\cdot 10^{30}$ & $350$\\
GM61        & $24$ & $74$ & $2^{61}-1$ & $1$ & $5.3\cdot 10^{36}$  & $415$ \\
GQ58.1      & $8  $ & $48 $ & $2^{29}(2^{29}-3)$ & $1$ & $\geq 2.8\cdot 10^{17}$ & $200$ \\
GQ58.3      & $8  $ & $48 $ & $2^{29}(2^{29}-3)$ & $3$ & $\geq 2.8\cdot 10^{17}$ & $200$ \\
GQ58.4      & $8  $ & $48 $ & $2^{29}(2^{29}-3)$ & $4$ & $\geq 2.8\cdot 10^{17}$ & $200$ \\
LFSR113     &$-$&$-$&$-$&$-$& $1.0\cdot 10^{34}$   & $30$  \\
MRG32K3A    &$-$&$-$&$-$&$-$& $3.1\cdot 10^{57}$   & $45$  \\
MT19937     &$-$&$-$&$-$&$-$& $4.3\cdot 10^{6001}$ & $623$ \\
\hline
\end{tabular}
\end{table}
\begin{table}[p]
\caption{Statistical testing of the generators which skip ahead $2^n$ values at each step.
}
\label{TestSkipAhead}
\footnotesize

\hspace{-2.7cm}
\begin{tabular}{|l|c|c|c|c||l|c|c|c|c|}
\hline
Generator    &  $n$  &  Smallcrush & Crush & Bigcrush &     Generator &  $n$  &  Smallcrush & Crush & Bigcrush \\
             &       &  (15 tests) & (144 tests) & (160 tests) &      &       &  (15 tests) & (144 tests) & (160 tests) \\
\hline
GM19         &    0  &    0        &   0   &      0    &  GQ58.1       &    0  &    0        &   0   &     0     \\
GM19         &    7  &    0        &  0(1) &  13(23)   &  GQ58.1       &    8  &    0        &   0   &     0     \\
GM19         &   14  &    3        & 69(81)&  122(129) &  GQ58.1       &   16  &    0        &  0(1) &     0     \\
GM19         &   21  &    0        &  0(1) &     0     &  GQ58.1       &   24  &    0        &   0   &     0     \\
\hline
GM29         &    0  &    0        &   0   &     0     &  GQ58.3       &    0  &    0        &   0   &     0     \\
GM29         &    8  &    0        &   0   &     0(1)  &  GQ58.3       &    8  &    0        &   0   &    0(2)   \\
GM29         &   16  &    0        &  0(2) &     0(1)  &  GQ58.3       &   16  &    0(1)     &   0   &     0     \\
GM29         &   24  &    0        &   0   &     0     &  GQ58.3       &   24  &    0        &   0   &     0     \\
\hline
GM31         &    0  &    0        &   0   &     0     &  GQ58.4       &    0  &    0        &   0   &     0     \\
GM31         &    8  &    0        &  0(1) &    0(1)   &  GQ58.4       &    8  &    0        &   0   &     0     \\
GM31         &   16  &    0        &   0   &    0(2)   &  GQ58.4       &   16  &    0        &  0(1) &     0     \\
GM31         &   24  &    0        &  0(1) &    0(1)   &  GQ58.4       &   24  &    0        &  0(1) &     0     \\
\hline
GM55         &    0  &    0        &   0   &     0     &  LFSR113      &    0  &     0       &   6   &     6     \\
GM55         &   20  &    0        &   0   &     0     &  LFSR113      &   25  &     0       &   6   &     6     \\
GM55         &   40  &    0        &   0   &     0     &  LFSR113      &   50  &     0       &   6   &     6     \\
GM55         &   60  &    0        &   0   &     0     &  LFSR113      &   75  &     0       &  6(7) &    6(7)   \\
\hline
GM61         &    0  &    0        &   0   &     0     &  MT19937      &    0  &     0       &   2   &     2     \\
GM61         &   25  &    0(1)     &   0   &     0(1)  &  MT19937      &   19  &     0       &   2   &    2(4)   \\
GM61         &   50  &    0        &   0   &     0     &  MT19937      &  200  &     0       &  2(3) &    2(3)   \\
GM61         &   75  &    0        &  0(2) &     0     &  MT19937      &  440  &     0       &  2(4) &     2     \\
\hline
MRG32K3A     &    0  &     0       &   0   &     0     &  MRG32K3A     &   60  &     0       &   0   &    0      \\
MRG32K3A     &   30  &     0       &   0   &     0     &  MRG32K3A     &   90  &     0       &   0   &    0(2)   \\
\hline
\end{tabular}
\end{table}

In this work we will widely use the block splitting method
for generation of parallel streams of pseudorandom 
numbers~(see also Sec.~\ref{StreamsSec}).
In order to empirically test the reliability of this method 
for chosen generators and to estimate the possibility of
long range correlations which could influence the statistical 
independence of parallel streams created with the block 
splitting method, we have carried out the statistical testing
of the generators skipping ahead $2^n$ values
at each step.
Table~\ref{TestSkipAhead} presents the results of such statistical testing
for several values of $n$ for each generator.
The number of failed tests is presented in the same way
as in Table~\ref{rngprop}.
If the number of statistical tests with p-values outside the
interval $[10^{-3},1-10^{-3}]$ differs from the number of statistical tests with p-values outside the
interval $[10^{-10},1-10^{-10}]$, then it is displayed inside the parenthesis.
Such obtained p-values (which are located outside of the first interval
and inside the second interval and are considered in Table~\ref{TestSkipAhead})
were found to be only slightly smaller 
than $10^{-3}$ (or slightly larger than $1-10^{-3}$) for all generators except GM19.
Therefore, they may appear in the table as statistical
fluctuations due to a large number of tests and may not witness serious correlations.
The bad properties of GM19 for $n\geq 7$ are explained by a very small period length of GM19.
Indeed, the GM19 period length is $2.7\cdot 10^{11}$.
Because we skip ahead $2^7=128$ values at each step,
the period is insufficient for tests in Bigcrush test battery:
each of them use about $3.4\cdot 10^{9}$ pseudorandom numbers on the average.
The results in Table~\ref{TestSkipAhead} show, as expected,
that skipping ahead at each step does not
deteriorate the statistical properties if the period length is sufficiently large.
We note that for each of the chosen generators,
the re-ordered sequence according to the chosen skip-ahead
is the sequence of the same type~(see also Sec.~\ref{StreamsSec}).
This explains the reasonably good properties of the re-ordered sequences.

\section{Parallel streams of pseudorandom numbers}
\label{StreamsSec}

There are several requirements for a good RNG and its implementation
in a subroutine library. Among them are statistical robustness
(uniform distribution and independence of values at the output,
absence of correlations), unpredictability, long period, efficiency,
theoretical support (precise prediction of the important properties),
portability, repeatability, ability to jump ahead,
proper initialization~\cite{Knuth,Lecuyer94,Brent,CatMaps2006}.
One more requirement allowing to use a parallel system for Monte Carlo
calculations is the ability to generate parallel streams
of pseudorandom numbers. Streams of pseudorandom numbers should
be independent and uncorrelated, i.e., statistical independence
within each stream and between the streams is necessary.
According to the theoretical classification of the parallelization
techniques in~\cite{BaukeMertens}, known techniques are:
random seeding, parametrization, block splitting, leapfrog.
Among these methods, block splitting and leapfrog are
most reliable. They require ability to jump ahead in the output
sequence. PRAND library includes the abilities to jump ahead
inside RNG sequence and to initialize independent random number streams
with block splitting method for each of the RNGs.

\subsection{Jumping ahead and initialization of streams
for GM19, GM31, GM61, GM29, GM55, GQ58.1, GQ58.3, GQ58.4}

The main recurrence relation of the generator of this type is~\cite{CatMaps2006}
\be
x_i^{(n)}=kx_i^{(n-1)}-qx_i^{(n-2)}\mymod{g},\ i=0,1,\dots,s-1.
\label{rec1}
\ee
Suppose that there exist integers $n,k_n,q_n$ such that
for any positive integer $w$
\be
x^{(2n+w)}=k_nx^{(n+w)}-q_nx^{(w)}\mymod{g}.
\label{rec2}
\ee
For example, for $n=1,k_1=k,q_1=q$, relation (\ref{rec2})
reduces to (\ref{rec1}).
It follows from (\ref{rec2}) that
$x^{(4n)}=(k_n^2-2q_n)x^{(2n)}-q_n^2x^{(0)}\mymod{g}$.
Thus, the values of parameters
for relation (\ref{rec2}) can be calculated
for $n\in\N$ with
\be
k_1=k;\quad q_1=q;\quad
k_{2n}=k_n^2-2q_n\mymod{g};\quad
q_{2n}=q_n^2\mymod{g}.
\ee
This allows to quickly jump ahead $n=2^i$ pseudorandom
numbers using the relation~(\ref{rec2}). In order to efficiently jump ahead $n$
pseudorandom numbers for arbitrary $n$ one can
subsequently skip ahead blocks of length
$2^{i_0}, 2^{i_1}, \dots, 2^{i_m}$, where
the binary notation of value $n$ is
$n=2^{i_0}+2^{i_1}+\dots+2^{i_m}$.
Alternatively, one can use the relations
\be
k_0=2; k_1=k; k_{n+1}=kk_n-qk_{n-1}\mymod{g}; q_n=q^n\mymod{g},
\label{rec3}
\ee
which can be proved by induction
and allow to efficiently jump ahead in the output sequence
using the relation~(\ref{rec2}).
Table~\ref{StreamsTable} shows maximal number of sequences
and maximal length of each sequence for each function
initializing pseudorandom stream.
It follows from the results in table~\ref{StreamsTable}
that the PRAND library can be efficiently used
for Monte Carlo calculations on present
and future supercomputers.

\begin{table}
\caption{Initialization of pseudorandom streams for RNGs in PRAND library}
\label{StreamsTable}
\begin{tabular}{|l|c|c|}
\hline
Function initializing sequence     & Number of sequences & Maximal length     \\
\hline
\verb#gm19_init_sequence_#             &    $1000$           & $6\cdot 10^6$      \\
\verb#gm29_init_short_sequence_#       &    $10^8$           & $8\cdot 10^7$      \\
\verb#gm29_init_medium_sequence_#      &    $10^6$           & $8\cdot 10^9$      \\
\verb#gm29_init_long_sequence_#        &    $10^4$           & $8\cdot 10^{11}$   \\
\verb#gm31_init_short_sequence_#       &    $10^9$           & $8\cdot 10^7$      \\
\verb#gm31_init_medium_sequence_#      &    $10^7$           & $8\cdot 10^9$      \\
\verb#gm31_init_long_sequence_#        &    $10^5$           & $8\cdot 10^{11}$   \\
\verb#gm55_init_short_sequence_#       &    $10^{18}$        & $10^{10}$          \\
\verb#gm55_init_long_sequence_#        &    $4\cdot 10^9$    & $10^{20}$          \\
\verb#gq58x1_init_short_sequence_#     &    $10^8$           & $8\cdot 10^7$      \\
\verb#gq58x1_init_medium_sequence_#    &    $10^6$           & $8\cdot 10^9$      \\
\verb#gq58x1_init_long_sequence_#      &    $10^4$           & $8\cdot 10^{11}$   \\
\verb#gq58x3_init_short_sequence_#     &    $2\cdot 10^8$    & $8\cdot 10^7$      \\
\verb#gq58x3_init_medium_sequence_#    &    $2\cdot 10^6$    & $8\cdot 10^9$      \\
\verb#gq58x3_init_long_sequence_#      &    $2\cdot 10^4$    & $8\cdot 10^{11}$   \\
\verb#gq58x4_init_short_sequence_#     &    $3\cdot 10^8$    & $8\cdot 10^7$      \\
\verb#gq58x4_init_medium_sequence_#    &    $3\cdot 10^6$    & $8\cdot 10^9$      \\
\verb#gq58x4_init_long_sequence_#      &    $3\cdot 10^4$    & $8\cdot 10^{11}$   \\
\verb#gm61_init_sequence_#             &  $1.8\cdot 10^{19}$ & $10^{10}$          \\
\verb#gm61_init_long_sequence_#        &    $4\cdot 10^9$    & $3\cdot 10^{25}$   \\
\verb#lfsr113_init_sequence_#          &  $3.8\cdot 10^{18}$ & $10^{10}$          \\
\verb#lfsr113_init_long_sequence_#     &    $4\cdot 10^9$    & $10^{24}$          \\
\verb#mrg32k3a_init_sequence_#         &       $10^{19}$     & $10^{38}$          \\
\verb#mt19937_init_sequence_#          &       $10^{19}$     & $10^{130}$         \\
\hline
\end{tabular}
\end{table}

\subsection{Jumping ahead and initialization of streams
for MRG32K3A}

The algorithm of MRG32K3A is based on the following
recurrence relations~\cite{MRG32K3A}:
\begin{align}
\mathbf{X_{n+1}}&=A\,\mathbf{X_n}\mymod{m_1}, \nonumber \\
\mathbf{Y_{n+1}}&=B\,\mathbf{Y_n}\mymod{m_2}.
\end{align}
Here
\be
\mathbf{X_n}=\begin{pmatrix}
x_n\\ x_{n-1}\\x_{n-2}
\end{pmatrix},\ \
\mathbf{Y_n}=\begin{pmatrix}
y_n\\ y_{n-1}\\y_{n-2}
\end{pmatrix},\ \
A=\begin{pmatrix}
0&a&b\\
1&0&0\\
0&1&0
\end{pmatrix},\ \
B=\begin{pmatrix}
c&0&d\\
1&0&0\\
0&1&0
\end{pmatrix},
\ee
$a=1403580$, $b=-810728$, $c=527612$, $d=-1370589$,
$m_1=2^{32}-209$, $m_2=2^{32}-22853$. The output sequence
is $\{(x_n+y_n)\mymod{g}\}$.
In order to jump ahead a block of length $n$
it is sufficient to find $A^n\mymod{m_1}$
and $B^n\mymod{m_2}$. This can be done in $O(\log n)$
operations.
Table~\ref{StreamsTable} shows that the function
\verb#mrg32k3a_init_sequence_# from PRAND library
can initialize up to $10^{19}$ independent
streams of length up to $10^{38}$.

\subsection{Jumping ahead and initialization of streams
for LFSR113}

The generator LFSR113~\cite{LFSR113} is a combination of the four
shift register sequences. Each of the shift register sequences
is based on a recurrence relation~\cite{Golomb}
\be
x_n = (x_{n-p}+ x_{n-p+q})\mymod{2},
\label{lfsrRel1}
\ee
where the output sequence is defined with
\be
u_n=\sum_{i=1}^{32} x_{ns+i-1} 2^{-i}.
\label{lfsrOutput}
\ee
The parameters of the four shift registers of LFSR113 are:
1) $p=31, q=6, s=18;$
2) $p=29, q=2, s=2;$
3) $p=28, q=13, s=7;$
4) $p=25, q=3, s=13.$

Consider a shift register sequence with the recurrence relation (\ref{lfsrRel1}),
where $p,q,s\in\{1,2,\dots,31\}$, $q<p$ and $p>16$.
As was mentioned in~\cite{Golomb}, for $k=2^e$ the following relation holds:
\be
x_n=(x_{n-kp}+x_{n-kp+kq})\mymod{2}.
\label{lfsrRel2}
\ee
This can be proved by induction as follows.
For $e=0$ the relation (\ref{lfsrRel2})
reduces to (\ref{lfsrRel1}).
For arbitrary $e\in\N$ we have
$x_n=x_{n-2^ep}+x_{n-2^ep+2^eq}\mymod{2}=
(x_{n-2^{e+1}p}+x_{n-2^{e+1}p+2^eq})+(x_{n-2^{e+1}p+2^eq}+x_{n-2^{e+1}p+2^{e+1}q})\mymod{2}=
x_{n-2^{e+1}p}+x_{n-2^{e+1}p+2^{e+1}q}\mymod{2}$.

Suppose we have the initial values $x_0, x_1,\dots, x_{31}$.
Let us build a table of values, which rows are numerated starting with zero.
The zero row of the table contains $p$ values: $x_p, x_{p+1},\dots, x_{2p-1}$,
which are directly calculated from the initial values.
The $n$-th row of the table
contain the following $p$ values: $x_{2^np},x_{2^n(p+1)},\dots,x_{2^n(2p-1)}$.
Each value of the table is calculated from the previous values with
the relation $x_{2^n(p+l)}=x_{2^n l}+x_{2^n(l+q)}$, which follows directly
from (\ref{lfsrRel2}). We note that for $n\geq 5$ the table contains $x_{s\cdot 2^n}$.
Indeed, it follows from $p>16$ that $s<2p$. If
$j\in\{0,1,2,3,4,5\}$ is maximal integer such that $s\cdot 2^j<2p$,
then $s\cdot 2^j\in\{p,p+1,\dots,2p-1\}$, therefore,
$x_{s\cdot 2^n}$ belongs to $(n-j)$-th row of the table.
In order to jump ahead a block of length $2^n$,
one needs to calculate the 32 bits $x_{s\cdot 2^n}, x_{s\cdot 2^n+1},\dots, x_{s\cdot 2^n+31}$,
which are involved in $u_{2^n}$ in (\ref{lfsrOutput}) for each of the four
shift registers of LFSR113.
To this purpose, one constructs $128$ tables which
start from each of the $32$ initial bits for
each of the four shift registers.
The realization of this operation for GPU is much more efficient
than for CPU because
the 128 tables can be constructed independently using different threads.
In order to efficiently jump ahead $n$
pseudorandom numbers for arbitrary $n$ one can
subsequently skip ahead blocks of length
$2^{i_0}, 2^{i_1}, \dots, 2^{i_m}$, where
the binary notation of value $n$ is
$n=2^{i_0}+2^{i_1}+\dots+2^{i_m}$.
Table~\ref{StreamsTable} shows maximal number of sequences
and maximal length of each sequence for each function
initializing pseudorandom stream.

\subsection{Jumping ahead and initialization of streams
for MT19937}
\label{MTJumpAhead}

The general algorithm for MT19937 is presented in~\cite{MT19937}.
The algorithm has a linear structure and can be written down as
$\mathbf{Y_{n+1}}=A\mathbf{Y_n}\mymod{2}$, where $\mathbf{Y_n}$
is a generator state. Therefore, the simplest algorithm
to jump ahead can be reduced to obtaining the matrix $A^n\mymod{2}$.
However, the size of matrix $A$ is $19937\times 19937$, therefore,
calculation of $A^n$ would be extremely slow and would require
considerable amount of memory. Much more efficient algorithm was suggested
in~\cite{Haramoto} and is based on the polynomial arithmetic in
the field $\F_2$. It is shown in~\cite{Haramoto} that for any $v\in\N$
the following relation holds:
\be
A^v\mathbf{Y_0}=g_v(A)\mathbf{Y_0}=a_k\mathbf{Y_{k-1}}+
a_{k-1}\mathbf{Y_{k-2}}+\dots+a_2\mathbf{Y_1}+a_1\mathbf{Y_0},
\label{g_v}
\ee
where $k=19937$,
$g_v(x)=a_kx^{k-1}+\dots+a_2x+a_1$ is the polynomial in the field $\F_2$,
and the coefficients $a_i\in\{0,1\}$, $i=1,2,\dots,k$, depend on $v$.
The algorithm for calculation of $g_v$ for arbitrary $v$ is presented
in~\cite{Haramoto}. This calculation is a consecutive calculation,
where massive parallelism of GPU could not be useful.
The calculation is relatively slow: it takes about several milliseconds
of modern CPU in order to calculate coefficients of polynomial $g_v(x)$
for a single value of $v$.
This time is sufficient to generate about 13 millions of pseudorandom numbers.
Therefore, this calculation should not be carried out right away.
As opposed to methods used in other software, in PRAND
the coefficients are already calculated for the following values of $v$:
$v=2^n, n=0,1,\dots,511$ and $v=n\cdot 2^i$, $n=15,16,\dots,24$,
$i=0,1,\dots,127$. Coefficients of each polynomial require $2496$
bytes in memory, therefore, less than $5$ megabytes are needed
in order to store all the calculated coefficients, which
is not a problem for modern GPUs.
In order to jump ahead a block of length $v$, it is sufficient
to calculate the sum (\ref{g_v}). The realization of this operation
for GPU is most efficient. If the coefficients of $g_v$ are not
immediately stored for a particular value of $v$, then jumping
ahead a block of length $v$ reduces to jumping ahead blocks of
lengths $2^{i_1},\dots,2^{i_l}$, where the binary notation
of value $v$ is $v=2^{i_1}+\dots+2^{i_l}$.
Table~\ref{StreamsTable} shows that the function
\verb#mt19937_init_sequence_#
can initialize up to $10^{19}$ independent
streams of length up to $10^{130}$.

\section{Multi-threaded generation of pseudorandom numbers with GPGPU}
\label{multiSec}

The PRAND library includes abilities of multi-threaded generation
of pseudorandom numbers with GPGPU for each of the RNGs.
The function call interface is described in Section~\ref{interface}.
In particular, the function \verb#rng_generate_gpu_array_#,
where \verb#rng# should be replaced with name of a generator,
fills in an array in GPU memory with pseudorandom numbers.
The parameters of this function are: the initial state of a
generator, the output array and its size.
The function \verb#rng_generate_array_#,
where \verb#rng# should be replaced with name of a generator,
generates pseudorandom numbers in GPU memory
and then copies them to an array in CPU memory.
To this purpose, an array is divided into sections,
and each section is filled independently using
one or several GPU threads. Prior to filling the
section with pseudorandom numbers, jumping ahead to
a starting point of the section takes place.

\begin{table}
\caption{Multi-threaded generation of pseudorandom numbers with GPGPU.}
\label{ArrayThreadsTable}
\begin{tabular}{|l|c|c|c|c|}
\hline
Generator & Number & Number of & Number of     & Number of \\
          & of     & threads   & threads per   & array sections \\
          & blocks & per block & array section &                \\
\hline
GM19     &  512  & 128 & 32  & 2048\\
GM29     &  512  & 128 & 32  & 2048\\
GM31     &  512  & 128 & 32  & 2048\\
GM55     &  128  & 128 & 8   & 2048\\
GM61     &  512  & 128 & 32  & 2048\\
GQ58.1   &  512  & 128 & 32  & 2048\\
GQ58.3   &  128  & 192 & 12  & 2048\\
GQ58.4   &  128  & 128 & 8   & 2048\\
LFSR113  &  512  & 128  & 128 & 512\\
MRG32K3A &  64   & 1024 & 1   & 65536\\
MT19937  &  64   & 227  & 227 & 64\\
\hline
\end{tabular}
\end{table}

Table~\ref{ArrayThreadsTable} shows the number
of GPU blocks and GPU threads used for
multi-threaded generation for each generator.
Also, table~\ref{ArrayThreadsTable}
shows the number of GPU threads used
in order to fill in a single array section.
The last column in Table~\ref{ArrayThreadsTable}
presents the number of array sections, which
can be calculated from the previous columns.
For example, for GQ58.3 the multi-threaded generation
employs ${128\times 192}\div{12}=2048$ array sections.

The function \verb#rng_generate_gpu_array_#,
where \verb#rng# should be replaced with
the name of the generator, requires that the
array size is divisible by the number of array
sections, which is shown in the last column in
Table~\ref{ArrayThreadsTable}. The particular
function \verb#mt19937_generate_gpu_array_#,
which fills in an array in GPU memory
with MT19937 pseudorandom numbers,
requires that the array size is a power of two.
The function \verb#rng_generate_array_#,
where \verb#rng# should be replaced with the name
of the generator, does not have any requirements
of this kind. If the array size is not
divisible by the number of array sections,
this function generates in GPU a slightly
larger amount of pseudorandom numbers, which
is divisible by the number of array sections,
and then copies only the required amount
of pseudorandom numbers to CPU memory.

\section{Function call interface}
\label{interface}
We briefly describe here the programming interface of PRAND library.
For additional information, the reader is encouraged to
check the content of the package directory \verb#examples#,
which contains numerous examples, and the content of the
package directory \verb#include#, which contains header files.
Below in this section we assume that
\verb#rng# should be replaced with the name
of a particular generator.

The PRAND library provides CPU as well as GPU interfaces.
Calling functions from device code
requires that all the data corresponding to the parameters,
such as generator's state, are kept in the device memory. 
Using the functions in host code requires that all the
data corresponding to the parameters are kept in the host memory. 
The only exception is multi-threaded generation
(see Sect.~\ref{MultiThreadedInterface} below).
It is possible to mix the CPU and GPU interfaces,
for example, advancing a generator first
on CPU and then -- from the same state -- on GPU.
In this case a user must ensure that the state 
and other necessary function parameters
are respectively copied between the device
memory and the host memory using \verb#cudaMemcpy#.

\subsection{Initialization}
Initialization can be performed both in host code
and in device code (i.e., both from CPU and from GPU)
with one of the following functions:

\noindent\begin{tabular}{l}
\verb#void rng_init_(rng_state* state);# \\
\verb#void rng_init_sequence_(rng_state* state, unsigned SequenceNumber);# \\
\verb#void rng_init_short_sequence_(rng_state* state, unsigned SequenceNumber);#  \\
\verb#void rng_init_medium_sequence_(rng_state* state, unsigned SequenceNumber);# \\
\verb#void rng_init_long_sequence_(rng_state* state, unsigned SequenceNumber);#
\end{tabular}

The function \verb#rng_init_# is supported for all of the generators.
The variable \verb#state# is a pointer to the structure of the
type \verb#rng_state#, which contains all the necessary
internal information for a particular generator and is initialized by the function \verb#rng_init_#.

Only those of the functions
\verb#rng_init_sequence_#,
\verb#rng_init_short_sequence_#,
\verb#rng_init_medium_sequence_#,
\verb#rng_init_long_sequence_# are supported, which are listed in table~\ref{StreamsTable}.
For the generators \verb#gm55#, \verb#gm61#, \verb#mrg32k3a#, \verb#lfsr113#, \verb#mt19937#
the maximal number of sequences exceeds $2^{32}$, in this case
the type of the parameter \verb#SequenceNumber# is~\verb#unsigned long long#.

\subsection{Jumping ahead}
Jumping ahead  can be performed both in host code and in device code
with the function\\ \verb#void rng_skipahead_(rng_state* state, unsigned long long offset)#.
Depending on the particular generator, there could be more parameters for the offset,
for example, the function \verb#gm55_skipahead_# contains three parameters:
\verb#state#, \verb#offset64# and \verb#offset0#; the function \verb#mrg32k3a_skipahead_#
contains four parameters: \verb#state#,
\verb#offset128#, \verb#offset64#
and \verb#offset0#. The header files contain particular list of parameters for each case.

\subsection{Single-threaded generation}
Single-threaded generation can be performed both in host code and in device code
with one of the following functions:

\noindent\begin{tabular}{l}
\verb#unsigned int rng_generate_(rng_state* state);# \\
\verb#float rng_generate_uniform_float_(rng_state* state);#
\end{tabular}

The function \verb#rng_generate_# generates a pseudorandom $32$-bit unsigned integer,
while the function \verb#rng_generate_uniform_float_# generates a pseudorandom
real number uniformly distributed in the interval $[0,1)$.

\subsection{Generation with SSE}
Generation with SSE can be performed only in host code with the following function:\\
\verb#unsigned int rng_sse_generate_(rng_sse_state* state);#\\
The parameter of type~\verb#rng_sse_state# can be obtained from a variable of
the type~\verb#rng_state# with the following function:\\
\verb#void rng_get_sse_state_(rng_state* state,rng_sse_state* sse_state);#

\subsection{Multi-threaded generation}
\label{MultiThreadedInterface}

Multi-threaded generation can be performed only in host code with one of the following functions:\\
\verb#void rng_generate_array_(rng_state* state, unsigned int* out, long length);# \\
\verb#void rng_generate_gpu_array_(rng_state* state, unsigned int* out, long length);#\\
\verb#void rng_generate_gpu_array_float_(rng_state* state, float* out, long length);#\\
\verb#void rng_generate_gpu_array_double_(rng_state* state, double* out, long length);#

The host function \verb#rng_generate_array_# fills in the array named 
\verb#"out"# in the host memory, while the host functions 
\verb#rng_generate_gpu_array_#, \verb#rng_generate_gpu_array_float_#,
\verb#rng_generate_gpu_array_double_# 
fill in the array named \verb#"out"# in device memory, and, therefore,
require that the array is located in device memory.
At the same time, the parameters \verb#state# and \verb#length# 
should be kept in the host memory for each of the four functions.
Multi-threaded generation is described in detail in Section~\ref{multiSec},
where one can find, in particular, the additional requirements 
on the size of the array named \verb#"out"# for the functions
\verb#rng_generate_gpu_array_#, \verb#rng_generate_gpu_array_float_#,
\verb#rng_generate_gpu_array_double_#.

\subsection{Header files}

In order to use a functionality of a particular generator,
a single corresponding header file should be included in a program,
for example: \verb#include<gm55.h>#. In this case, one can use
any host functions described above, including host functions 
which carry out multi-threaded generation with GPU.

In order to use device functions from used-defined device code, 
it is necessary to include 
the PRAND source file corresponding to a particular RNG, for example: 
\verb#include"/home/user/prand/source/_gm55.cu"#.
In this case it is not necessary to specify the whole PRAND library
for the linking stage of compilation by CUDA compiler.

\subsection{Fortran compatibility}

Fortran compatibility has been tested. The examples of using the PRAND library
from Fortran are included in the~\verb#examples# directory.

\subsection{Function call interface for MRG32K3A}

Function call interface for MRG32K3A is slightly different.
Prior to using the MRG32K3A capabilities from the device code and
prior to using the multi-threaded generation of MRG32K3A pseudorandom numbers,
the following additional initialization should be performed:
\verb#mrg32k3a_init_device_consts_();#\\
After using the features under discussion, the following call should be performed:
\verb#mrg32k32_free_device_consts_();#\\
In order to perform jumping ahead from device code, one should use the following function\\
\verb#void mrg32k3a_dev_skipahead_(mrg32k3a_state* state,#
\verb#unsigned long#
\verb# long offset128,#
\verb#unsigned long long offset64,#
\verb#unsigned long long offset0);#\\
In order to initialize parallel stream from device code, one should use
the function:\\
\verb#void mrg32k3a_dev_init_sequence_(mrg32k3a_state* state,#
\verb#unsigned#
\verb#long long SequenceNumber);#\\
Initialization of parallel streams from host code and jumping ahead from host code
can be performed in a usual way, as described in previous subsections.

\subsection{Function call interface for MT19937}

Function call interface for MT19937 is slightly different.
Prior to using the MT19937 capabilities from the device code and
prior to using the multi-threaded generation of MT19937 pseudorandom numbers,
the following additional initialization should be performed:
\verb#mt19937_init_device_consts_();#\\
After using the features under discussion, the following call should be performed:
\verb#mt19937_free_device_consts_();#\\
In order to initialize parallel stream from device code, one should use
the following function:\\
\verb#void mt19937_dev_init_sequence_(mt19937_state* state,#
\verb#unsigned#
\verb#long#
\verb#long SequenceNumber);#\\
In order to initialize parallel stream from host code, the function
\verb#mt19937_init_sequence# with the same parameters should be used.
In order to perform jumping ahead from device code, one should use the function\\
\verb#void mt19937_dev_skipahead_(mt19937_state* state, unsigned long long#
\verb#a,unsigned b);#\\
This function skips ahead $N=a\cdot 2^b$ numbers, where $N<2^{512}$.
In order to perform jumping ahead from host code, the function
\verb#mt19937_skipahead_# with the same parameters as in \verb#mt19937_dev_skipahead#
should be used.

\subsection{Examples}

The library contains a lot of examples which illustrate usage of most of its features.
The examples can be found in the~\verb#examples# directory.

\section{Performance tests}
\label{PerformanceSec}

\begin{table}
\caption{Speed of generation for different realizations. 
CPU: Intel Xeon X5670; GPU:  Nvidia Fermi C2050; 
OS: Linux SLES 11 SP1 / CentOS 5.5;
Compiler: CUDA 5.0; Optimization: -O2.}
\label{PerfTable}
\begin{tabular}{|l|c|c|c|c|}
\hline
          & CPU,         & CPU, & GPU,            & GPU,            \\
Generator & ANSI C       & SSE & single-threaded  & multithreaded   \\
          & (numbers/sec) & (numbers/sec) & (numbers/sec) & (numbers/sec) \\
\hline
GM19     & $7.3\cdot 10^6$ & $3.3\cdot 10^7$  & $1.6\cdot 10^5$  & $2.6\cdot 10^8$ \\
GM29     & $8.5\cdot 10^6$ & $3.6\cdot 10^7$  & $2.1\cdot 10^5$  & $2.5\cdot 10^8$ \\
GM31     & $9.1\cdot 10^6$ & $3.1\cdot 10^7$  & $7.3\cdot 10^4$  & $2.1\cdot 10^8$ \\
GM55     & $2.1\cdot 10^7$ & $4.6\cdot 10^7$  & $1.9\cdot 10^5$  & $5.6\cdot 10^8$ \\
GM61     & $4.2\cdot 10^6$ & $8.9\cdot 10^6$  & $1.5\cdot 10^4$  & $1.6\cdot 10^8$ \\
GQ58.1   & $7.8\cdot 10^6$ & $1.3\cdot 10^7$  & $2.2\cdot 10^4$  & $1.9\cdot 10^8$ \\
GQ58.3   & $1.5\cdot 10^7$ & $2.4\cdot 10^7$  & $1.3\cdot 10^5$  & $3.8\cdot 10^8$ \\
GQ58.4   & $2.1\cdot 10^7$ & $4.0\cdot 10^7$  & $1.9\cdot 10^5$  & $5.4\cdot 10^8$ \\
LFSR113  & $2.2\cdot 10^8$ & $1.6\cdot 10^8$  & $1.2\cdot 10^6$  & $1.3\cdot 10^8$ \\
MRG32K3A & $5.0\cdot 10^7$ & $1.5\cdot 10^8$  & $7.3\cdot 10^5$  & $2.1\cdot 10^9$ \\
MT19937  & $2.0\cdot 10^8$ & $2.5\cdot 10^8$  & $6.1\cdot 10^5$  & $2.9\cdot 10^9$ \\
\hline
\end{tabular}
\end{table}

\begin{table}
\caption{Comparing performance of PRAND, NAG Numerical routines for GPUs and Nvidia cuRand.
The task fills $100$ times an array of $n=2^{29}$ double-precision floating point
numbers in GPU memory with pseudorandom numbers.
CPU: Intel Xeon E5630; GPU: Nvidia Tesla X2070; OS: CentOS 6.1;
Compiler: CUDA 5.0; Optimization: -O2.}
\label{CompareTable}
\begin{tabular}{|l|c|c|}
\hline
Library & Time (sec) & Time (sec) \\
& MRG32K3A & MT19937/MTGP32 \\
\hline
CURAND & 9.6 & 16.3 \\
NAG & 10.5 & 13.7 \\
PRAND & 14.5 & 13.4 \\
\hline
\end{tabular}
\end{table}

\begin{figure}
\caption{A simple program in CUDA C, which tests the performance of MTGP32 
from Nvidia cuRand library: 
the task fills $100$ times an array of $n=2^{29}$ single-precision floating point
numbers in GPU memory with pseudorandom numbers.}
\label{TestingMTGP}
\begin{verbatim}
#include <stdio.h>
#include <curand.h>
int main(){
  size_t n = 536870912;
  clock_t start=clock(); int i;
  curandGenerator_t gen;
  float *devData; float f;
  cudaMalloc((void **)&devData, n * sizeof(float));
  curandCreateGenerator(&gen,CURAND_RNG_PSEUDO_MTGP32);  
  curandSetPseudoRandomGeneratorSeed(gen, 1234ULL);
  for(i=0;i<100;i++) curandGenerateUniform(gen, devData, n);
  printf("time1 = %f seconds\n",(float)(clock()-start)/CLOCKS_PER_SEC);
  cudaMemcpy(&f, devData, sizeof(float),cudaMemcpyDeviceToHost);
  printf("time2 = %f seconds\n",(float)(clock()-start)/CLOCKS_PER_SEC);
  curandDestroyGenerator(gen);
  cudaFree(devData);
  printf("time3 = %f seconds\n",(float)(clock()-start)/CLOCKS_PER_SEC);
  return 0;
}
\end{verbatim}
\end{figure}

Table~\ref{PerfTable} shows performance for different realizations.
The testing was done on the following platform:
CPU: Intel Xeon X5670; GPU:  Nvidia Fermi C2050; 
OS: Linux SLES 11 SP1 / CentOS 5.5;
Compiler: CUDA 5.0; Optimization: -O2.
The result shows that multi-threaded generation with GPU
could be several thousands times faster than single-threaded generation.

Table~\ref{CompareTable} compares performance
of PRAND, NAG Numerical routines for GPUs and Nvidia cuRand.
The table contains timings for the task
which fills $100$ times an array of $n=2^{29}$ double-precision floating point
numbers in GPU memory with pseudorandom numbers.
We note that Nvidia cuRand does not have
realization of MT19937, in this case the comparison
with performance of MTGP32 is performed.

The benchmarks in Tables~\ref{PerfTable} and \ref{CompareTable}
seem to yield multithreaded GPU results significantly
slower than those reported in~\cite{Manssen} which indicate,
e.g., $18\cdot 10^9$ numbers/sec for MTGP32 in Nvidia cuRand
library. Figure~\ref{TestingMTGP} shows a simple program in C language
which obtains the performance of MTGP32 generator from Nvidia cuRand
library. The task fills $100$ times an array of $n=2^{29}$ single-precision floating point
numbers in GPU memory with pseudorandom numbers, and then fixes the
time point \verb#time1#. After that the task copies a {\em single} number
from the device memory to the host memory, and then fixes the time point \verb#time2#.
Finally, the task destroys the generator, frees all the previously allocated memory
and fixes the time point \verb#time3#.
The result of this program on the platform containing
Intel Xeon X5670 (CPU) and Nvidia Fermi C2050 (GPU) is:
\verb#time1 = 3.36 sec.#,
\verb#time2 = 22.18 sec.#,
\verb#time3 = 22.18 sec.#
The result on the platform containing
Intel Xeon E5630 (CPU) and Nvidia Tesla X2070 (GPU) is:
\verb#time1 = 1.64 sec.#,
\verb#time2 = 20.75 sec.#,
\verb#time3 = 20.75 sec.#
The program shows the same result if the \verb#cudaMemcpy# function 
call is replaced, for example, by call of \verb#cudaThreadSynchronize# function.

Using the value of \verb#time1# one obtains $16\cdot 10^9$ numbers/sec
for the first platform and $33\cdot 10^9$ numbers/sec for the second platform.
However, the value of \verb#time2# or \verb#time3# results only in
$2.4\cdot 10^9$ numbers/sec for the first platform and
$2.6\cdot 10^9$ numbers/sec for the second platform.
Obviously, at the time point of \verb#time1# the
calculations in GPU are not yet completed.
Therefore, only \verb#time2# or \verb#time3# describe the actual
performance of the generator. This is a possible reason of
the difference in the performances of multithreaded GPU generation reported 
in the present work and in~\cite{Manssen}.

\section{Conclusion}

PRAND can be considered as GPU accelerated implementation
of \-RNGSSELIB library~\cite{RNGSSELIB1,RNGSSELIB2}.
PRAND library contains realization of a number of modern and reliable generators:
\verb#MT19937#, \verb#MRG32K3A#, \verb#LFSR113#,
\verb#GM19#, \verb#GM29#, \verb#GM31#, \verb#GM61#,
\verb#GM55#, \verb#GQ58.1#, \verb#GQ58.3# and \verb#GQ58.4#.
The library contains: single-threaded and multi-threaded realizations for GPU,
single-threaded realizations for CPU, realizations for CPU based on SSE command set.
Abilities to jump ahead inside RNG sequence and to initialize independent random number streams
with block splitting method for each of the RNGs are included in the library.
Fortran compatibility has been tested and the examples of using the PRAND library
from Fortran are included. In addition to detailed description of the library,
this paper contains in Sec.~\ref{softwareSec} a brief review of other existing software
for random number generation, and also in Sec.~\ref{PropGen} the paper
contains analysis of properties of many known RNGs.

This work was partially supported by the Russian Foundation for Basic Research
project Nos 12-07-13121 and 13-07-00570,
by the Supercomputing Center of Lomonosov Moscow State University~\cite{Lomonosov}
and by a Marie Curie International Research Staff Exchange Scheme Fellowship
within the European Community's Seventh Framework Programme.
The authors acknowledge The Numerical Algorithms Group Ltd
for the access to the RNG NAG software with purpose of testing and comparison.

\end{document}